\documentclass[a4paper,11pt]{article}
\usepackage{axodraw4j}
\usepackage{jheppub} % for details on the use of the package, please
\usepackage[T1]{fontenc} % if needed
\title{\boldmath \rat{} vertices for the effective ggH theory}

\author{Ben Page and Roberto Pittau}
\affiliation{Departamento de F\'{\i}sica Te\'orica y del Cosmos and CAFPE,\\
Campus Fuentenueva s. n., Universidad de Granada 
E-18071 Granada, Spain}
\emailAdd{page@ugr.es,pittau@ugr.es}

\abstract{We list all possible \rat{} Feynman rules needed in NLO computations involving couplings of Higgs and gluons mediated by an infinitely heavy top loop.
They provide the rational contribution generated by the $(d-4)$-dimensional part of the amplitude, paving the way for four-dimensional automatic NLO methods in Higgs phenomenology.}

\begin{document} 
\newcommand{\bqa}{\begin{eqnarray}}
\newcommand{\eqa}{\end{eqnarray}}
\newcommand{\rat}{${\rm R_2}$}
\newcommand{\qbar}{\bar q}
\newcommand{\nl}{\nonumber \\}
\newcommand{\als}{\alpha_S}

\maketitle
\flushbottom
\section{Introduction}
Automation of NLO calculations is an important issue in modern high energy particle physics. The complexity of the processes one has to deal with forbids a case by case approach, forcing the use of automatic tools~\cite{Ossola:2007ax,Giele:2008bc,vanHameren:2009dr,Hirschi:2011pa,Frederix:2011ss,Bevilacqua:2011xh,Bern:2013gka,Cullen:2011ac,Mastrolia:2010nb}. 

In the last few years, a lot of progress has been achieved and, among the various available 
techniques~\cite{Bern:1994cg,Britto:2004nc,Forde:2007mi,Berger:2008sj,Ellis:2009zw}, the Ossola-Papadopoulos-Pittau (OPP) method~\cite{Ossola:2006us} appears to be a rather convenient one. OPP is a procedure which allows one to numerically extract, from dimensionally regulated one-loop amplitudes, the Cut-Constructible part plus a rational piece - called ${\rm R_1}$ - strictly linked to the four-dimensional numerator of the Feynman diagrams. A second contribution to the rational part, dubbed \rat{}, cannot be obtained numerically in four-dimensions, and must be provided externally. A particularly convenient way to account for \rat{} is via effective Feynman rules. This is because the \rat{} contribution disappears when a sufficiently large number of particles undergo the scattering and, in renormalizable theories, only up to four-particle-vertices have to be considered.

The \rat{} vertices for QCD~\cite{Draggiotis:2009yb}, the Electroweak Model~\cite{Garzelli:2009is,Garzelli:2010qm,Shao:2011tg} and MSSM~\cite{Shao:2012ja} have already been presented in the literature. In this paper, we list the \rat{} Feynman rules generated by the interaction of one Higgs field $H$ with two, three and four gluons - mediated by an infinitely heavy top loop - encoded in the effective Lagrangian~\cite{Shifman:1979eb,Dawson:1991au}
\bqa
\label{efflagr}
{\cal L}_{\rm eff} &=& -\frac{1}{4} A H G^{a}_{\mu \nu} G^{a,\mu \nu}\,, 
\eqa
where
\bqa
\label{eq:A}
A &=& \frac{\als}{3\pi v}\left(1+\frac{11}{4}\frac{\als}{\pi} \right)
\eqa
and $v$ is the vacuum expectation value, $v^2= (G_F \sqrt{2})^{-1}$.

There is a very important physical motivation that led us to study this problem: the main Higgs production mechanism at the LHC is via gluon fusion~\cite{Aad:2012tfa,Chatrchyan:2012ufa}, and considering NLO corrections to processes generated by the effective operator in eq.~(\ref{efflagr}) provides an easy way to model, at two-loop precision, Higgs + $n$ jets final states for jets with 
${\rm p_{T}}$ not much larger than the top mass~\cite{DelDuca:2001eu,DelDuca:2001fn,Campanario:2013mga}.
  The $Hjj$ and $Hjjj$ signatures have been recently considered~\cite{vanDeurzen:2013rv,Cullen:2013saa}, in the context of a $d$-dimensional integrand reduction technique, within the GoSam~\cite{Cullen:2011ac}/Sherpa~\cite{Gleisberg:2008ta} framework. In this paper we provide the only missing ingredient needed by 4-dimensional NLO approaches, such as OPP, Open Loops~\cite{Cascioli:2011va} or FDR~\cite{Pittau:2012zd}, to attack the problem in a fully automatic fashion.

In the next section we review the basics of \rat{}. 
In section~\ref{vertices} we list the contributing \rat{} vertices and
section~\ref{conc} contains our conclusions.

\section{\rat{} in a nutshell}
\label{r2}
The full theory of \rat{} can be found in~\cite{Ossola:2008xq,Pittau:2011qp}. For the purposes of this paper, it is enough to recall that, starting from a one-loop amplitude computed in $d$ dimensions, \rat{} is generated by the explicit splits
\bqa
\label{splits}
d &=& 4 + \epsilon\,, \nl
\qbar^2 &=& q^2 + \tilde{q}^2 
\eqa
in the numerators of the contributing Feynman diagrams, where $\qbar^2$ is the
$d$-dimensional integration momentum squared, $q^2$ its four-dimensional part and $\tilde{q}^2$ the difference between the two. 
In renormalizable theories, in the limit $\epsilon \to 0$, the $\epsilon$ and $\tilde{q}^2$ pieces generate constant terms in one particle irreducible Green's functions up to four legs.
The effective operator in eq.~(\ref{efflagr}) has dimension five and so a non-vanishing contribution can survive in up to five-leg vertices. 

In Dimensional Reduction like schemes~\cite{Siegel:1979wq}, one is free to set $d= 4$ in eq.~(\ref{splits}) right from the beginning, while the $\tilde{q}^2$ part is always necessary in order to keep gauge invariance. The transition rules between Dimensional Reduction and the usual 't Hooft-Veltman Dimensional Regularization~\cite{'tHooft:1972fi}, in which $d = 4 + \epsilon$, are well 
known~\cite{signer,Pittau:2011qp}. To allow the reading of our rules in both schemes we introduced a parameter $\lambda_{\rm HV}$ in our formulae, such that
\bqa
d &=& 4 + \lambda_{\rm HV}\,\epsilon\,.
\eqa
Thus, Dimensional Reduction corresponds to $\lambda_{\rm HV}= 0$ and
Dimensional Regularization to $\lambda_{\rm HV}= 1$.

\section{The \rat{} vertices generated by the $GGH$ operator}
\label{vertices}
By power counting, a non-zero \rat{} contribution is present only in the following five interactions
\bqa
Hgg,~~Hggg,~~Hgggg,~~Hq \bar q,~~Hq \bar q g\,.    
\eqa 
We generated and computed, with the help of {\tt QGRAF}~\cite{Nogueira:1991ex} 
and {\tt FORM}~\cite{Vermaseren:2000nd}, all possible contributing 
diagrams~\footnote{We used the QCD Feynman rules in appendix B of~\cite{Draggiotis:2009yb}, together with the effective Higgs-gluons couplings listed in~\cite{Kauffman:1996ix}.} from which we extracted \rat{} via eq.~(\ref{splits}).  
Two independent calculations were performed, with slightly different 
strategies. In the first computation a Feynman parametrization was applied before eq.~(\ref{splits}); in the second, eq.~(\ref{splits}) was used to classify independent integrals, which were computed at a later stage of the calculation. Both procedures gave the same expressions, listed in figure~\ref{fig:fig1}, which represent the main result of this paper.

The $Hggg$ vertex is fully proportional to the QCD three gluon vertex
\begin{align}
    V&^{\mu_1 \mu_2 \mu_3}_{m_1 m_2 m_3}(p_1, p_2, p_3)
    =  \notag \\
    &+g_s f_{m_1 m_2 m_3} \Big[g^{\mu_1 \mu_2} (p_2 - p_1)^{\mu_3} +
      g^{\mu_2 \mu_3} (p_3 - p_2)^{\mu_1} 
     +g^{\mu_3 \mu_1}(p_1 -p_3)^{\mu_2}\Big]\,,
\end{align}
while the tensor involved in the Higgs/4-Gluons vertex reads
\begin{align}
     X&^{\mu_1 \mu_2 \mu_3  \mu_4}_{m_1 m_2 m_3 m_4}
    = \notag \\ 
    &+ {\rm Tr} (T^{m_1}T^{m_2}T^{m_3}T^{m_4}) \Big[ 
                                + 21\, g^{\mu_1 \mu_2} g^{\mu_3 \mu_4} 
                                - 41\, g^{\mu_1 \mu_3} g^{\mu_2 \mu_4} 
                                + 21\, g^{\mu_1 \mu_4} g^{\mu_2 \mu_3} 
                          \Big] \notag \\
    &+ {\rm Tr} (T^{m_1}T^{m_2}T^{m_4}T^{m_3}) \Big[ 
                                + 21\, g^{\mu_1 \mu_2} g^{\mu_3 \mu_4} 
                                + 21\, g^{\mu_1 \mu_3} g^{\mu_2 \mu_4} 
                                - 41\, g^{\mu_1 \mu_4} g^{\mu_2 \mu_3} 
                          \Big] \notag \\
    &+ {\rm Tr} (T^{m_1}T^{m_3}T^{m_2}T^{m_4}) \Big[
                                - 41\, g^{\mu_1 \mu_2} g^{\mu_3 \mu_4} 
                                + 21\, g^{\mu_1 \mu_3} g^{\mu_2 \mu_4} 
                                + 21\, g^{\mu_1 \mu_4} g^{\mu_2 \mu_3} 
                          \Big]\,,
\end{align}
where the color structure is compactly expressed in terms of traces of color matrices $T^a$ in the {\em adjoint} representation. The translation to traces of Gell-Mann matrices 
$t^a$ in the fundamental representation of $SU(N_c)$ is given by the formula
\bqa
{\rm Tr} (T^aT^bT^cT^d)= 
N_c\,\left[
{\rm Tr}(t^at^bt^ct^d)+{\rm Tr}(t^dt^ct^bt^a) 
\right] + \frac{\delta_{ab}\delta_{cd}+
                \delta_{ac}\delta_{bd}+
                \delta_{ad}\delta_{bc}}{2}\,.
\eqa

Any $\lambda_{\rm HV}$ contribution is entirely contained in the \rat{} part. However, in physical combinations no dependence on $\lambda_{\rm HV}$ remains. Naturally this means that the vertices involving only gluons are devoid of $\lambda_{\rm HV}$. Furthermore one can show that in the physical sum of diagrams, given by
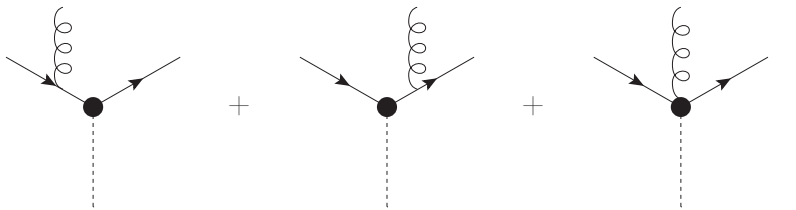
\begin{center}
\colorbox{white}{
  \begin{picture}(300,92.3) (0,0)
    \SetScale{0.75}
    \SetOffset(0,10) 
    \Line[arrow,arrowpos=0.5,arrowlength=5,arrowwidth=2,arrowinset=0.2](50,50)(93.3,75)
    \Line[arrow,arrowpos=0.5,arrowlength=5,arrowwidth=2,arrowinset=0.2](6.7,75)(50,50)
    \Vertex(50,50){5}
    \Line[dash,dashsize=2](50,0)(50,50)
    \Gluon(35,59)(35,100){4.2}{3}
    \Text(88,38)[l]{+}
    \SetOffset(110,10) 
    \Line[arrow,arrowpos=0.5,arrowlength=5,arrowwidth=2,arrowinset=0.2](50,50)(93.3,75)
    \Line[arrow,arrowpos=0.5,arrowlength=5,arrowwidth=2,arrowinset=0.2](6.7,75)(50,50)
    \Vertex(50,50){5}
    \Line[dash,dashsize=2](50,0)(50,50)
    \Gluon(65,59)(65,100){4.2}{3}
    \Text(88,38)[l]{+}
    \SetOffset(220,10) 
    \Line[arrow,arrowpos=0.5,arrowlength=5,arrowwidth=2,arrowinset=0.2](50,50)(93.3,75)
    \Line[arrow,arrowpos=0.5,arrowlength=5,arrowwidth=2,arrowinset=0.2](6.7,75)(50,50)
    \Vertex(50,50){5}
    \Line[dash,dashsize=2](50,0)(50,50)
    \Gluon(50,54)(50,100){4.2}{3}
    \Text(80,35)[l]{,}
  \end{picture}
}
\end{center}
there is no dependence on $\lambda_{\rm HV}$.

Finally, it is important to realize that, being one particle irreducible, the vertices 
in figure~\ref{fig:fig1} are the building blocks from which the \rat{} contribution to {\em any} amplitude involving one coupling of order $A$~\footnote{See eq.~(\ref{eq:A}).} and {\em any } number of additional jets can be derived.

\begin{figure}
\begin{center}
\colorbox{white}{
  \begin{picture}(300,92.3) (0,0)
    \SetScale{0.75}
    \SetOffset(-27,0) 
    \SetWidth{1.0}
    \SetColor{Black}
    %Slants
    % 50(1+sin(60))~ 93.3
    % 50(1+cos(60))= 75
    \Gluon
         (50,50)(93.3,75){4.7}{5}
    % 80(1-sin(60))~ 10.7
    \Gluon
         (6.7,75)(50,50){4.7}{5}
    %Vertical
    \Line[dash,dashsize=2](50,0)(50,50)
    \Vertex(50,50){5}
    \Text(105, 29)[lb]{{$\displaystyle{ =\frac{iAg_s^2N_c\delta^{m_1 m_2}}{384\pi^2}\,
                    \Big\{p_1^{\mu_1} p_2^{\mu_2} 
                    + 89\, p_1^{\mu_2} p_2^{\mu_1} 
                    + 14\,(p_1^{\mu_1} p_1^{\mu_2} + p_2^{\mu_1} p_2^{\mu_2})}$}}
    \Text(192, 0)[lb]{{$\displaystyle{- \left[17\,(p_1^2 + p_2^2) + 93\, (p_1 \cdot p_2)\right]
                       g^{\mu_1 \mu_2}\Big\}}$}}
    \Text(-27, 65)[lb]{$p_1, \,  \mu_1, \,  m_1$}
    \Text(50, 65)[lb]{$p_2, \, \mu_2, \, m_2$}
  \end{picture}
}
\end{center}

\begin{center}
\colorbox{white}{
  \begin{picture}(300,92.3) (0,0)
    \SetScale{0.75}
    \SetOffset(-27,0) 
    \SetWidth{1.0}
    \SetColor{Black}
    %Horizontals
    \Gluon(50,50)(100,50){4.7}{5}
    \Gluon(0,50)(50,50){4.7}{5}
    %Vertical
    \Gluon(50,50)(50,100){4.7}{5}
    \Line[dash,dashsize=2](50,0)(50,50)
    \Vertex(50,50){5}
    \Text(105, 32)[lb]{{$\displaystyle{ =-\frac{15Ag_s^2N_c}{128\pi^2}\,
                        V^{\mu_1 \mu_2 \mu_3}_{m_1 m_2 m_3}(p_1, p_2, p_3)}$}}
    \Text(16, 82)[lb] {$p_2, \, \mu_2, \, m_2$}
    \Text(-27, 17)[lb]{$p_1, \, \mu_1, \, m_1$}
    \Text(50, 17)[lb] {$p_3, \, \mu_3, \, m_3$}
  \end{picture}
}
\end{center}

\begin{center}
\colorbox{white}{
  \begin{picture}(300,92.3) (0,0)
    \SetScale{0.75}
    \SetOffset(-27,0) 
    \SetWidth{1.0}
    \SetColor{Black}
    %Slants
    % 50(1+sin(36))~ 79.4
    % 50(1+cos(36))~ 90.4
    \Gluon(79.4,90.4)(50,50){4.7}{5}
    % 50(1-sin(18))~ 34.6
    % 50(1+cos(18))~ 97.5
    \Gluon(50,50)(97.5,34.6){4.7}{5}
    % 50(1-cos(18))~ 2.4
    \Gluon(2.4,34.6)(50,50){4.7}{5}
    % 50(1-sin(36))~ 20.6
    \Gluon(50,50)(20.6,90.4){4.7}{5}
    
    %Nudged to look right
    \Line[dash,dashsize=2](50,0)(50,50)

    \Vertex(50,50){5}

    \Text(105, 30)[lb]{{$\displaystyle{ = \frac{iAg_s^4}{128\pi^2}\,
                       X^{\mu_1 \mu_2 \mu_3 \mu_4}_{m_1 m_2 m_3 m_4}}$}}
    \Text(-22, 75)[lb]{$\mu_2, \, m_2$}
    \Text(91.5, 75)[rb] {$\mu_3, \, m_3$}
    \Text(91.5, 9)[rb]  {$\mu_4, \, m_4$}
    \Text(-22, 9)[lb] {$\mu_1, \, m_1$}
  \end{picture}
}
\end{center}

\begin{center}
\colorbox{white}{
  \begin{picture}(300,92.3) (0,0)
    \SetScale{0.75}
    \SetOffset(-22,0) 
    \SetWidth{1.0}
    \SetColor{Black}
    %Slants
    % 50(1+sin(60))~ 93.3
    % 50(1+cos(60))= 75
    \Line[arrow,arrowpos=0.5,arrowlength=5,arrowwidth=2,arrowinset=0.2](50,50)(93.3,75)
    % 50(1-sin(60))~ 6.7
    \Line[arrow,arrowpos=0.5,arrowlength=5,arrowwidth=2,arrowinset=0.2](6.7,75)(50,50)
    %Vertical
    \Vertex(50,50){5}
    \Line[dash,dashsize=2](50,0)(50,50)
    \Text(105, 30)[lb]{{$\displaystyle{ = \frac{iAg_s^2}{32\pi^2}
    \left(\frac{N_c^2-1}{2 N_c}\right)\delta^{j_1 j_2} \lambda_{\rm HV} (\rlap/p_1 -
    \rlap/p_2) }$}}
    \Text(-22,62)[lb]{$p_1,\, j_1$}
    \Text(66,62)[lb]{$p_2, \, j_2$}
  \end{picture}
}
\end{center}

\begin{center}
\colorbox{white}{
  \begin{picture}(300,92.3) (0,0)
    \SetScale{0.75}
    \SetOffset(-22,0) 
    \SetWidth{1.0}
    \SetColor{Black}
    %Horizontals
    \Line[arrow,arrowpos=0.5,arrowlength=5,arrowwidth=2,arrowinset=0.2](0,50)(50,50)
    \Gluon(50,50)(100,50){4.7}{5}
    %Verticals
    \Line[arrow,arrowpos=0.5,arrowlength=5,arrowwidth=2,arrowinset=0.2](50,50)(50,100)
    \Vertex(50,50){5}
    \Line[dash,dashsize=2](50,0)(50,50)
    \Text(0, 42)[lb]{$j_1$}
    \Text(25, 68)[lb]{$j_2$}
    \Text(74, 25)[lb]{$\mu,\, m$}
    \Text(105,28)[lb]{{$\displaystyle{ =\frac{iAg_s^3}{64\pi^2}\gamma_{\mu}
    t_{j_2 j_1}^m
    \left[\frac{2\lambda_{\rm HV}+1}{N_c} - (2\lambda_{\rm HV}+3) N_c\right]}$}}
  \end{picture}
}
\end{center}
\caption{\label{fig:fig1} The \rat{} vertices generated by the effective Lagrangian in eq.~(\ref{efflagr}). All momenta are incoming, $N_c$ is the number of colors and $\lambda_{\rm HV}= 1~(\lambda_{\rm HV}=0)$ in Dimensional Regularization (Reduction). Quarks are massless and the dashed line represents the Higgs field. The tensors $V^{\mu_1 \mu_2 \mu_3}_{m_1 m_2 m_3}(p_1, p_2, p_3)$ and $X^{\mu_1 \mu_2 \mu_3 \mu_4}_{m_1 m_2 m_3 m_4}$ are defined in the text.}
\end{figure}
\section{Conclusions}
\label{conc}
We have presented the complete set of effective \rat{} Feynman rules generated by 
the dimension five operator $GGH$. They encode the missing analytical information needed to apply four-dimensional automatic integrand reduction techniques to Higgs physics. 

Our result can be used to study both production and decay processes involving one Higgs particle and any number of jets. This is particularly useful for Higgs phenomenology at the LHC, where, due to the large amount of available energy, the Higgs particle is very often produced in association with a large number of jets.

\acknowledgments
This work was performed in the framework of the ERC grant 291377, ``LHCtheory: Theoretical predictions and analyses of LHC physics: advancing the precision frontier''. We also thank the support of the MICINN project FPA2011-22398 (LHC@NLO) and the Junta de Andaluc\'{i}a project P10-FQM-6552.

\bibliography{r2ggh}{}
\bibliographystyle{JHEP}
\end{document}